\documentclass[aps,prl,superscriptaddress,reprint]{revtex4-1}
\usepackage{blindtext}
\usepackage{amsmath}
\usepackage{subcaption}
\usepackage{graphicx,float}
\usepackage{mathptmx}

\begin{document}
\title{Regulating Anderson Localization with Structural Defect Disorder}
\author{Mouyang Cheng}
\affiliation{School of Physics, Peking University, Beijing 100871, People’s Republic of China}

\author{Haoxiang Chen}
\affiliation{School of Physics, Peking University, Beijing 100871, People’s Republic of China}

\author{Ji Chen}
\email{ji.chen@pku.edu.cn}
\affiliation{School of Physics, Peking University, Beijing 100871, People’s Republic of China}
\affiliation{Interdisciplinary Institute of Light-Element Quantum Materials and Research Center for Light-Element Advanced Materials, Peking University, Beijing 100871, People's Republic of China}
\affiliation{Frontiers Science Center for Nano-Optoelectronics, Peking University, Beijing 100871, People's Republic of China}
\affiliation{Collaborative Innovation Center of Quantum Matter, Beijing 100871, People’s Republic of China}

\begin{abstract}

Localization due to disorder has been one of the most intriguing theoretical concepts evolved in condensed matter.
Here, we expand the theory of localization by considering two types of disorder at the same time, namely the original Anderson's disorder and the structural defect disorder, which has been suggested to be a key component in recently discovered two-dimensional amorphous materials.
While increasing the degree of both disorders could induce localization of wavefunction in real space, we find that a small degree of structural defect disorder can significantly enhance the localization.
As the degree of structural defect disorder increases, localized states quickly appear within the extended phase to enter a broad crossover region with mixed phases. 
Full localization occurs when structural defect disorder reaches about 10 percent, even without considering the Anderson type disorder.
Our theoretical model provides a comprehensive understanding of localization in two-dimensional amorphous materials and highlights the promising tunability of their transport properties.

\end{abstract}

\maketitle

%\section{I. Introduction}

The phenomenon that sufficient disorder results in space localization of wavefunction originates from the model of Anderson (1958)\cite{r1}.
A delocalized-localized transition was found at a critical order parameter value $W/V$, where $W$ reflects the disorder of on-site energy, and $V$ is the hopping integral between neighboring sites.
Anderson's study inspired many theoretical works in the last century \cite{Lee1985}, including the variable-range hopping\cite{r8,r9} and the scaling theory of localization \cite{r10,r11,r12}. 
While analytical and numerical approaches fail to reach a consensus on the exact critical transition value of $W/V$, they all manage to show the same qualitative picture\cite{r13,r14,r15,r16}. 
Besides, a large number of modified Anderson's models\cite{web1,web2,web3,web4,web5,web6,web7} show the universality of wave localization in disordered systems; 
experiments in microwaves, acoustics, band gap materials and cold atoms all agree with the prediction\cite{r2,r3,r4,r5,r6,r7} that the transport behaviour exhibits a drastic change due to wavefunction localization. 
Disorder induced localization also plays important roles in the search for topological quantum states \cite{li_topological_2009,groth_theory_2009,song_dependence_2012}.
These works highlight that the localization phenomenon has a general physical picture and broad impact in various research fields. 

Although the original Anderson's localization model is widely applied to interpret the delocalized-localized transition and the change of transport properties in non-crystalline materials, recent works have indicated that it is not always sufficient for realistic disordered materials.
For example, two-dimensional (2D) amorphous monolayer carbon has been recently achieved in experiments, providing an ideal platform to examine structural-property relationship with unprecedented precision \cite{kotakoski_from_2011,eder_journey_2014,joo2017realization,toh_synthesis_2020}. 
In particular, Tian et al.\cite{tian_disorder-tuned_2023} reported a tuning of electrical conductivity up to 9 orders of magnitude via controlling the degree-of-disorder in amorphous monolayer carbon.
A variable hopping model was employed in the same work to explain the conductivity change, where a link was made between the electric conductivity and the structural disorder. 
The structural disorders are associated with the so-called medium range order and the structural defects, including structural holes and contaminated regions. 
While the physics of medium range order can boil down to the original Anderson's model, the role of the structural defects can not be captured by the original model and the existing extensions.
In addition to monolayer carbon, structural defects can also be created in a large number of 2D materials using different experimental techniques, providing opportunity to engineer their properties with structural defect disorders \cite{Zhao_two_2019,hong2020ultralow,shin_preferential_2021,zhang_structure_2022}.

In this work, we propose a new extension of the Anderson localization model by considering simultaneous two different mechanisms of disorder on the 2D lattice. 
Two order parameters are introduced, one corresponds to the original Anderson's disorder ($W/V$) and one characterizes the structural defect disorder. 
We then perform systematic numerical calculations and analyses, which show that both disorders can induce wavefunction localization.
A diagram of localization is given, where the cooperation and competition of two localization mechanisms are fully revealed.
The electrical conductivity is calculated using the Kubo-Greenwood formula\cite{Kubo1957,Greenwood1958}, which further illustrates that both types of disorder have significant impacts on the electron transport.

%\section{II. Results}

%\subsection{Model and two degrees of disorder}

We start from the non-interacting tight-binding Hamiltonian on a 2D square lattice with a single electron band,
$$
\begin{aligned}
\hat H &=\sum_{j,k}\left(V_{x,jk}\hat{a}_{j+1,k}^\dagger \hat{a}_{j,k}+V_{y,jk}\hat{a}_{j,k+1}^\dagger \hat{a}_{j,k}\right)+\text{h.c.} \\
&+\sum_{j,k}E_{jk}\hat{a}_{j,k}^\dagger \hat{a}_{j,k}
\end{aligned}
$$
where $V_{x,jk}$ and $V_{y,jk}$ stand for the hopping energy from site $(j,k)$, 
$E_{jk}$ is potential energy of site $(j,k)$.
$\hat{a}_{j,k}^\dagger$ and $\hat{a}_{j,k}$ are one electron creation and annihilation operators at site $(j,k)$. 
A discussion about the relation between our model and real material is detailed in the supplementary information (SI) section I.
Diagonalization of the corresponding hopping matrix results in the Hamiltonian's eigenstates.
In the Anderson localization model, disorder is described by a random distribution of the on-site energy between $[\epsilon-W/2,\epsilon+W/2]$.
When neglecting the off-diagonal disorder, namely choosing a constant hopping term $V$, a dimensionless order parameter ($W/V$) is used to characterize the degree of disorder. 
Existing theoretical and computational works on realistic amorphous materials have focused on disorders caused structural and topological distortions, establishing a good knowledge of the structural-property relationships \cite{kapko_electronic_2010, van_tuan_insulating_2012, thapa_ab_2022, zhang_structure_2022}. 
These types of disorders can be converted to the description with $W/V$.

\begin{figure}[H]
  \centering
  \includegraphics[width=0.5\textwidth]{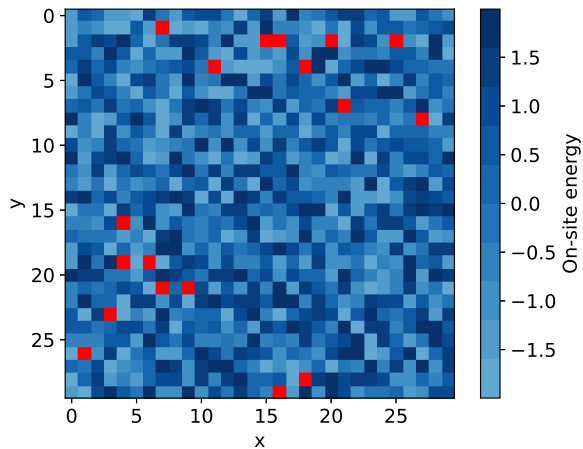}
  \caption{A potential energy map on a 30$\times$30 square lattice with two types of disorder. The two order parameters are $W/V=4$ and $\lambda=0.02$. The red isolated patches indicate random distributed forbidden sites with $E_m=1000$. The blue colored patches correspond to the disorder from random distribution of the on-site energy.}
\end{figure}

Inspired by the structural defects in 2D amorphous materials, such as atomic holes and contaminated regions, we introduce a new parameter to describe the disorder, namely the ratio of forbidden sites for electrons ($\lambda $). 
To model this new disorder, we randomly choose some of the sites on the square lattice, and set their on-site potential as $E_m$, which is significantly larger than the magnitude of the on-site energy and the hopping kinetic energy. 
The effect is obvious that the extremely large on-site potential creates a huge eigenvalue, making the corresponding site physically forbidden for electrons occupation. 
Thus, the direct outcome is that these high energy sites are forbidden for electron hopping, which may also contribute to the localization of wavefunction.

Fig. 1 shows a map of the on-site energy on a 30$\times$30 square lattice, illustrating the coexistence of the two types of disorder.
The red patches corresponds to the forbidden sites and the on-site energy fluctuation due to Anderson disorder are shown with blue colors.
We expect both types of disorder make contribution to localization of wavefunction and reduce the electrical conductivity. 
In the below, we discuss numerical results performed on a $30\times 30$ square lattice, where $W/V$ varies from 0 to 8 and $\lambda $ is within 0.25
\footnote{Our model has a constraint that $\lambda<0.5$. This is because as $\lambda>0.5$, the wavefunction starts to delocalize as $\lambda$ continues to rise. This can be easily shown at $\lambda=1$, where the Hamiltonian is simply
$
\hat{H}(\lambda=1)=E_m\hat{I}+\hat{H}(\lambda=0)
$
resulting in the same eigenstates as $\hat{H}(\lambda=0)$, which returns to the $\lambda=0$ case.
}.
The Fermi energy is set as $E_F=0$ and the energy of forbidden sites are set as $E_m=1000$.
More details, convergence tests, additional results and extended analyses are discussed in the SI.

%\section{III.Results}

\begin{figure}[H]
    \includegraphics[width=0.5\textwidth]{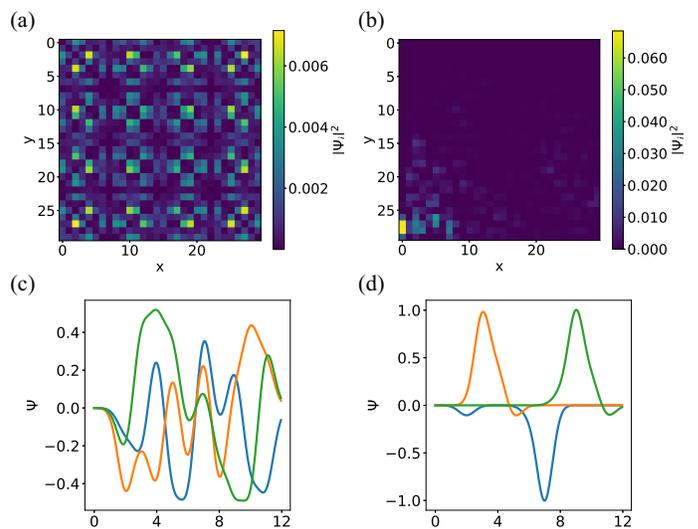}
    \caption{
    Visualization of the extended and localized states.
    (a,b) Occupation number of a randomly selected state on the two dimensional lattice from an extended system ($W/V=0$; $\lambda=0$) and a localized system ($W/V=0$; $\lambda=0.2$).
    (c,d) Three wavefunctions plotted along one lattice dimension. The wavefunctions are real due to the Hermitian matrix. 
    }
    \label{fig.2}
\end{figure}

To have a direct view of how wavefunction localizes, we calculate the occupation number $|\psi_i|^2$ for each site $i$. 
Fig. \ref{fig.2}a shows an example of the extended state, where many sites have non-zero occupations. 
For a localized state, the electron only occupies one site as displayed in Fig. \ref{fig.2}b.
A larger set of visualization diagrams are presented in SI.II.
In Fig. \ref{fig.2}c,d we show the wavefunction of three states in the extended system and the localized system along one lattice dimension, respectively. 
For an extended system, wavefunction of many states are delocalized as in Fig. \ref{fig.2}c, while for an localized system every state has its electron occupying one site.
In addition to the extended phase and the localized phase, we also identify the existence of a broad transition regime where mixed phases may occur.

To quantify the localization behavior, we calculate the normalized participation ratio (NPR)\cite{r17,Roy2021}, which is defined as
$$
\mathrm{NPR}_i=\left(L^2 \sum_{n=1}^{N^2}\left|\phi_i^n\right|^4\right)^{-1}
$$
where $N^2$ sites are considered, $L=Na$ is size of the system, $a$ is the lattice constant and $\phi_i^n$ is the component of the $i$-th eigen-function projected on $n$-th site.
For 2D systems, $\text{NPR}$ should satisfy the following conditions. 
$$
\text{NPR} \sim \begin{cases}\xi^{-2} & \text { for extended states }  
\\ L^{-2} & \text { for localized states. }\end{cases}
$$
$\xi$ is a constant characterizing the localization length of the corresponding states.
Therefore, for large $L$, NPR remains finite for extended states and vanishes for localized states. 

\begin{figure}[H]
\centering
    \includegraphics[width=0.5\textwidth]{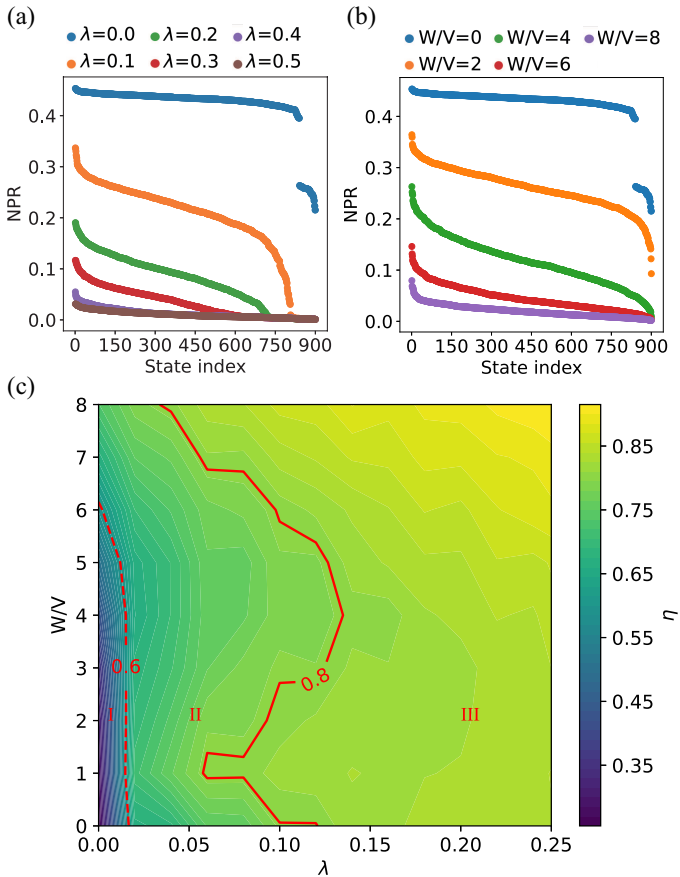}
    \caption{Quantification of localization under different order parameters. 
    (a,b) NPR values plotted for different eigenstates under different order parameters. Finite and vanishing NPR correspond to the extended and localized states respectively. (a) NPR for $W/V=0$ and $\lambda=0,0.1,0.2,0.3,0.4,0.5$ (lines from top to bottom). (b) NPR for $\lambda=0$ and $W/V=0,2,4,6,8$ (lines from top to bottom).
    (c) The localization contour diagram plotting $\eta$ as a function of the order parameters $W/V$ and $\lambda$. The red lines indicate the two boundaries ($\eta=0.6$ and $\eta=0.8$) separating three phase regimes: I. the extended phase, II. the crossover regime with mixed phases and III. the localized phase.  
    }
    \label{fig.localize}
\end{figure}

Fig. \ref{fig.localize}a,b plots NPR values of all eigenstates under $\lambda=0\text{-}0.5, W/V=0$ and $W/V=0\text{-}8,\lambda=0$.
Each NPR curve represents one set of parameters.
According to our design of the Hamiltonian, as $\lambda$ increases only $\lambda N^2$ sites should have vanishing NPR, and one expects a discontinuous function of the NPR curve plotted as a function of the state index.
However, Fig. \ref{fig.localize}a shows the curves drop to zero  
continuously apart from the fully extended system (blue line).
The overall non-vanishing NPR of the other states also decays to zero as $\lambda$ increases.  
At around $\lambda=0.4$, the whole NPR curve is reduced to $\sim L^2$, where a full localization of the whole system occurs.
These observations show that (i) structural defect disorder can induce a mixture of localized and extended states simultaneously in the system; (ii) a small portion of structural defects can cause global localization of the whole system.
In contrast, the original Anderson type disorder $W/V$ induces a smoother transition from extended phase at $W/V=0$ to around $W/V=6$ where most of the states have vanishing NPR.

If we take these two order parameters together into consideration, the system consists of a rich variety of phases. 
To further classify them, we define a new parameter $\eta=\frac{\langle \text{NPR} \rangle(N/2)}{4\langle \text{NPR} \rangle(N)} $, where
$\langle \text{NPR} \rangle(N)$ is the average value of all eigenstates' NPR for an $N \times N$ lattice.
The definition of $\eta$ is inspired by the fact that for the fully localized phase NPR $\sim N^{-2}$, while for the extended phase NPR of all eigenstates is independent of $N$. 
With such a definition, we have $\eta=0.25$ for pure extended states, and  $\eta=1$ for pure localized states. 
Fig. \ref{fig.localize}c is a colored contour diagram of $\eta$ as a function of the two order parameters $W/V$ and $\lambda$.

We classify the extended, crossover and localized phases revealed above as I, II and III, respectively.
As a result there are two steps in the localization transition, namely the step from the extended phase to the crossover regime with mixed phases; and the step to the fully localized phase.
To this end, we choose two critical values for $\eta$, which are indicated as read lines on Fig. 3. For the boundary between the extended phase and the crossover regime, we recall that the original Anderson transition point is near $W/V \approx 6$\cite{mott2012electronic} and $\lambda=0$, so we take the contour line of $\eta$ which crosses this conventional transition point, i.e. $\eta_1=0.6$, to represent the boundary. 
The second boundary is set at $\eta_2=0.8$ because numerical calculations show that as $\lambda$ rises, $\eta$ drastically increases to $\eta=0.75 \sim 0.85$ and then slowly approaches $\eta=0.9 \sim 1$ for larger disorder (SI.III). We note that in real systems $\eta$ may not reach the ideal value 1 because of finite $N$ and incompleteness of localization. Therefore, $\eta_2=0.8$ can be considered as the boundary entering the fully localized phase.

It is quite clear that the three phase regions are well separated by the variation of $\lambda $. 
Tuning the ratio of randomly distributed forbidden sites within a range of approximately 10\% can already achieve the transition from the extended phase to the fully localized phase. 
This is not affected significantly by the choice of the other parameter $W/V$ given that the value is within a reasonable range.
These results suggest that via introducing only a few percent of randomly distributed forbidden sites, one can achieve a strong enhancement of localization, providing a powerful solution to tuning the transport properties of materials and other systems.
In the SI.IX we show results calculated on a honeycomb lattice, where the same conclusion can be reached.

Another interesting observation is that the two types of disorder are not only trivially additive to each other, but cooperation and competition might exist.
With a finite $\lambda $, contour lines of $\eta$ extends towards the left as $W/V$ increases from 0 to 1, then rightwards until $W/V$ increases to 4.
The change of the direction of contour lines mean the localization firstly increases and is then suppressed before $W/V = 4$.
The reason for such an anomaly is further discussed in SI.V.
The key point is that the introduction of a small portion of forbidden sites can significantly alter the behavior of Anderson localization.

To further understand how the two types of disorder affect transport property, we calculate the conductivity $\sigma$ using the Kubo-Greenwood formula\cite{Kubo1957,Greenwood1958}. 
$$
\begin{aligned}
\sigma &=\lim_{\omega \to 0}\left(\pi / \omega L^{2}\right) \int d E^{\prime} \sum_{\alpha, \beta}\sum_{j}\left|\langle\alpha|J_x(j)| \beta\rangle\right|^{2} \\ 
&\times \delta\left(E^{\prime}+\omega-E_{\beta}\right) \delta\left(E^{\prime}-E_{\alpha}\right)
\end{aligned}
$$
where $L$ is the length of the system, $|\alpha\rangle$, $|\beta\rangle$ are eigenstates of the Hamiltonian, and correspond to eigenvalues $E_{\alpha}<E<E_{\beta}$. For a current flows in the $x-$ direction, $J_x(j)$ is the current operator, and is defined as
$$
J_x(j)=i e \sum_{k} V_{x,jk}\left(a_{j, k}{ }^{\dagger} a_{j-1, k}-\text { h.c. }\right)
$$
More computational details are discussed in SI.VI.

\begin{figure}[H]
    \includegraphics[width=0.45\textwidth]{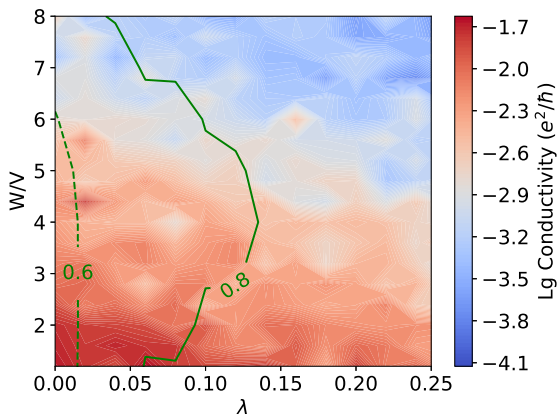}
    \caption{Conductivity diagram as a function of the order parameters $W/V$ and $\lambda$. For better visualization, the $W/V<1.2$ part is not shown because the decay near $W\sim 0+$ is too rapid. Green lines show the two localization boundaries from Fig. 3 ($\eta_{1,2}=0.6,0.8$).}
    \label{fig.conduct}
\end{figure}

Fig. \ref{fig.conduct} plots the conductivity for different parameters of $W/V$ and $\lambda $. 
The two critical contour lines with values $\eta_1$ and $\eta_2$ separating regions of different degrees of localization are also shown. 
Although it is expected that the increase of both disorders can lead to the reduction of conductivity, 
it is interesting to see that the effects of disorder on the conductivity do not follow the exactly same trend as the localization. 
It suggests that different from the conventional Anderson localization, a metal-insulator transition is no longer necessarily synchronized with the extended-localized transition.
Only a small portion of the forbidden sites are needed to cause the localization, but to induce a reduction of the conductivity a quite large number of forbidden sites are needed.
This is reasonable because localized electrons can still contribute to conductivity via hopping mechanisms, and only when all the hopping paths are blocked by enough number of forbidden sites, the conductivity can be strongly suppressed. 
From the perspective of the density of states (DOS) as displayed in SI.VIII, DOS near the Fermi energy decreases more rapidly as a function of $W/V$ than $\lambda$.

It is worth noting that in the studies of Tian et al. the insulating samples have many structural holes and contaminated sites \cite{tian_disorder-tuned_2023}, corresponding to a $\lambda >0.5$ and $W/V <3 $ (SI.X).
In that regime a variable range hopping model is more suitable and can describe the conductivity behavior very well. 
In the SI.IX we also show numerical results and analyses on the honeycomb lattice, which resembles the true structure of amorphous monolayer carbon. 
The enhancement of localization due to structural defect disorder is also observed, supporting the main conclusion of this study.
However, we shall note that the honeycomb lattice involves a new degree of complexity due to the semi-metallic nature.
One can notice the anomalous behavior in the weak localization regime and other differences in other parameter regimes.
While some behaviors can be straightforwardly understood by the semi-metallic band structure \cite{carva2010defect,anomaly_carbon}, further theoretical works are desired to reveal more details about the impact of structural defect disorder \cite{khveshchenko_electron_2006,onoda_localization_2007}.

In the future, when the quality of 2D amorphous materials are improved to a regime where the number of forbidden sites is within 20\%, we expect the conductivity to follow the current model.
Even in this regime, one can achieve a tuning of the conductivity by as much as 2 orders of magnitude.
In contrast, when $\lambda $ is small, the Anderson disorder can suppress the conductivity more effectively, where the conductivity reduces by at most 7 magnitudes when $W/V$ increases beyond 6. 
A similar behavior of both disorders is that the decay of conductivity mainly occurs at weak localization, at namely small $W/V$ and $\lambda $ (SI.VII).

%\section{IV. Conclusion}
To conclude, in this work we introduce an extension of the Anderson's localization model with randomly distributed forbidden sites, inspired by the structural defect disorder in 2D amorphous materials such as structural holes and contaminating sites.
Numerical calculations and analyses show that the structural defect disorder can enhance the localization, driving the extended phase first to a mixed phase then to a fully localized phase.
Importantly, a small portion of forbidden sites results in the level of localization equivalent to a very large degree of Anderson's disorder.
Conductivity calculations also suggest that a collective tuning of the two types of disorder can provide a more versatile control of the electrical conductivity of 2D materials.
Overall, our model provides a more comprehensive understanding of the localization phenomenon and electronic transport in 2D systems, which shall motivate a series of further investigations, for example, the impact of forbidden sites on other transport properties \cite{zhang_thermal_2022}; the possibility of novel emergent phenomena with the interplay of the two types of disorder \cite{agarwala_topological_2017,wang_structural_2022}; 
and the role of structural defect disorder in three dimensional disordered systems \cite{guo_topological_2010,ying_anderson_2016}.

\section{Acknowledgements}
The authors thank Huifeng Tian and Lei Liu for providing original experimental data and helpful discussions. We also thank Huaqing Huang for helpful discussions.
This work was supported by the National Natural Science Foundation of China under Grant No. 92165101, the National Key R\&D Program of China under Grant No. 2021YFA1400500, the Strategic Priority Research Program of Chinese Academy of Sciences under Grant No. XDB33000000, and the Beijing Natural Science Foundation (No. JQ22001). We are grateful for computational resources supported by High-performance Computing Platform of Peking University.

%%%reference
\bibliographystyle{apsrev4-1} % Tell bibtex which bibliography style to use
\bibliography{ref.bib} % Tell bibtex which .bib file to use (this one is some example file in TexLive's file tree)

\begin{thebibliography}{54}%
\makeatletter
\providecommand \@ifxundefined [1]{%
 \@ifx{#1\undefined}
}%
\providecommand \@ifnum [1]{%
 \ifnum #1\expandafter \@firstoftwo
 \else \expandafter \@secondoftwo
 \fi
}%
\providecommand \@ifx [1]{%
 \ifx #1\expandafter \@firstoftwo
 \else \expandafter \@secondoftwo
 \fi
}%
\providecommand \natexlab [1]{#1}%
\providecommand \enquote  [1]{``#1''}%
\providecommand \bibnamefont  [1]{#1}%
\providecommand \bibfnamefont [1]{#1}%
\providecommand \citenamefont [1]{#1}%
\providecommand \href@noop [0]{\@secondoftwo}%
\providecommand \href [0]{\begingroup \@sanitize@url \@href}%
\providecommand \@href[1]{\@@startlink{#1}\@@href}%
\providecommand \@@href[1]{\endgroup#1\@@endlink}%
\providecommand \@sanitize@url [0]{\catcode `\\12\catcode `\$12\catcode
  `\&12\catcode `\#12\catcode `\^12\catcode `\_12\catcode `\%12\relax}%
\providecommand \@@startlink[1]{}%
\providecommand \@@endlink[0]{}%
\providecommand \url  [0]{\begingroup\@sanitize@url \@url }%
\providecommand \@url [1]{\endgroup\@href {#1}{\urlprefix }}%
\providecommand \urlprefix  [0]{URL }%
\providecommand \Eprint [0]{\href }%
\providecommand \doibase [0]{http://dx.doi.org/}%
\providecommand \selectlanguage [0]{\@gobble}%
\providecommand \bibinfo  [0]{\@secondoftwo}%
\providecommand \bibfield  [0]{\@secondoftwo}%
\providecommand \translation [1]{[#1]}%
\providecommand \BibitemOpen [0]{}%
\providecommand \bibitemStop [0]{}%
\providecommand \bibitemNoStop [0]{.\EOS\space}%
\providecommand \EOS [0]{\spacefactor3000\relax}%
\providecommand \BibitemShut  [1]{\csname bibitem#1\endcsname}%
\let\auto@bib@innerbib\@empty
%</preamble>
\bibitem [{\citenamefont {Anderson}(1958)}]{r1}%
  \BibitemOpen
  \bibfield  {author} {\bibinfo {author} {\bibfnamefont {P.~W.}\ \bibnamefont
  {Anderson}},\ }\href {\doibase 10.1103/PhysRev.109.1492} {\bibfield
  {journal} {\bibinfo  {journal} {Physical Review}\ }\textbf {\bibinfo {volume}
  {109}},\ \bibinfo {pages} {1492} (\bibinfo {year} {1958})}\BibitemShut
  {NoStop}%
\bibitem [{\citenamefont {Lee}\ and\ \citenamefont
  {Ramakrishnan}(1985)}]{Lee1985}%
  \BibitemOpen
  \bibfield  {author} {\bibinfo {author} {\bibfnamefont {P.~A.}\ \bibnamefont
  {Lee}}\ and\ \bibinfo {author} {\bibfnamefont {T.~V.}\ \bibnamefont
  {Ramakrishnan}},\ }\href {\doibase 10.1103/RevModPhys.57.287} {\bibfield
  {journal} {\bibinfo  {journal} {Reviews of Modern Physics}\ }\textbf
  {\bibinfo {volume} {57}},\ \bibinfo {pages} {287} (\bibinfo {year}
  {1985})}\BibitemShut {NoStop}%
\bibitem [{\citenamefont {Mott}(1968{\natexlab{a}})}]{r8}%
  \BibitemOpen
  \bibfield  {author} {\bibinfo {author} {\bibfnamefont {N.~F.}\ \bibnamefont
  {Mott}},\ }\href {\doibase 10.1103/RevModPhys.40.677} {\bibfield  {journal}
  {\bibinfo  {journal} {Reviews of Modern Physics}\ }\textbf {\bibinfo {volume}
  {40}},\ \bibinfo {pages} {677} (\bibinfo {year}
  {1968}{\natexlab{a}})}\BibitemShut {NoStop}%
\bibitem [{\citenamefont {Mott}(1968{\natexlab{b}})}]{r9}%
  \BibitemOpen
  \bibfield  {author} {\bibinfo {author} {\bibfnamefont {N.~F.}\ \bibnamefont
  {Mott}},\ }\href {\doibase 10.1080/14786436808223200} {\bibfield  {journal}
  {\bibinfo  {journal} {Philosophical Magazine}\ }\textbf {\bibinfo {volume}
  {17}},\ \bibinfo {pages} {1259} (\bibinfo {year}
  {1968}{\natexlab{b}})}\BibitemShut {NoStop}%
\bibitem [{\citenamefont {Anderson}\ \emph {et~al.}(1980)\citenamefont
  {Anderson}, \citenamefont {Thouless}, \citenamefont {Abrahams},\ and\
  \citenamefont {Fisher}}]{r10}%
  \BibitemOpen
  \bibfield  {author} {\bibinfo {author} {\bibfnamefont {P.~W.}\ \bibnamefont
  {Anderson}}, \bibinfo {author} {\bibfnamefont {D.~J.}\ \bibnamefont
  {Thouless}}, \bibinfo {author} {\bibfnamefont {E.}~\bibnamefont {Abrahams}},
  \ and\ \bibinfo {author} {\bibfnamefont {D.~S.}\ \bibnamefont {Fisher}},\
  }\href {\doibase 10.1103/PhysRevB.22.3519} {\bibfield  {journal} {\bibinfo
  {journal} {Physical Review B}\ }\textbf {\bibinfo {volume} {22}},\ \bibinfo
  {pages} {3519} (\bibinfo {year} {1980})}\BibitemShut {NoStop}%
\bibitem [{\citenamefont {Abrahams}\ \emph {et~al.}(1979)\citenamefont
  {Abrahams}, \citenamefont {Anderson}, \citenamefont {Licciardello},\ and\
  \citenamefont {Ramakrishnan}}]{r11}%
  \BibitemOpen
  \bibfield  {author} {\bibinfo {author} {\bibfnamefont {E.}~\bibnamefont
  {Abrahams}}, \bibinfo {author} {\bibfnamefont {P.~W.}\ \bibnamefont
  {Anderson}}, \bibinfo {author} {\bibfnamefont {D.~C.}\ \bibnamefont
  {Licciardello}}, \ and\ \bibinfo {author} {\bibfnamefont {T.~V.}\
  \bibnamefont {Ramakrishnan}},\ }\href {\doibase 10.1103/PhysRevLett.42.673}
  {\bibfield  {journal} {\bibinfo  {journal} {Physical Review Letters}\
  }\textbf {\bibinfo {volume} {42}},\ \bibinfo {pages} {673} (\bibinfo {year}
  {1979})}\BibitemShut {NoStop}%
\bibitem [{\citenamefont {Lee}\ and\ \citenamefont {Fisher}(1981)}]{r12}%
  \BibitemOpen
  \bibfield  {author} {\bibinfo {author} {\bibfnamefont {P.~A.}\ \bibnamefont
  {Lee}}\ and\ \bibinfo {author} {\bibfnamefont {D.~S.}\ \bibnamefont
  {Fisher}},\ }\href {\doibase 10.1103/PhysRevLett.47.882} {\bibfield
  {journal} {\bibinfo  {journal} {Physical Review Letters}\ }\textbf {\bibinfo
  {volume} {47}},\ \bibinfo {pages} {882} (\bibinfo {year} {1981})}\BibitemShut
  {NoStop}%
\bibitem [{\citenamefont {Edwards}\ and\ \citenamefont {Thouless}(1972)}]{r13}%
  \BibitemOpen
  \bibfield  {author} {\bibinfo {author} {\bibfnamefont {J.~T.}\ \bibnamefont
  {Edwards}}\ and\ \bibinfo {author} {\bibfnamefont {D.~J.}\ \bibnamefont
  {Thouless}},\ }\href {\doibase 10.1088/0022-3719/5/8/007} {\bibfield
  {journal} {\bibinfo  {journal} {Journal of Physics C: Solid State Physics}\
  }\textbf {\bibinfo {volume} {5}},\ \bibinfo {pages} {807} (\bibinfo {year}
  {1972})}\BibitemShut {NoStop}%
\bibitem [{\citenamefont {Stein}\ and\ \citenamefont {Krey}(1980)}]{r14}%
  \BibitemOpen
  \bibfield  {author} {\bibinfo {author} {\bibfnamefont {J.}~\bibnamefont
  {Stein}}\ and\ \bibinfo {author} {\bibfnamefont {U.}~\bibnamefont {Krey}},\
  }\href {\doibase 10.1007/BF01325498} {\bibfield  {journal} {\bibinfo
  {journal} {Zeitschrift fur Physik B Condensed Matter and Quanta}\ }\textbf
  {\bibinfo {volume} {37}},\ \bibinfo {pages} {13} (\bibinfo {year}
  {1980})}\BibitemShut {NoStop}%
\bibitem [{\citenamefont {Thouless}(1974)}]{r15}%
  \BibitemOpen
  \bibfield  {author} {\bibinfo {author} {\bibfnamefont {D.}~\bibnamefont
  {Thouless}},\ }\href {\doibase 10.1016/0370-1573(74)90029-5} {\bibfield
  {journal} {\bibinfo  {journal} {Physics Reports}\ }\textbf {\bibinfo {volume}
  {13}},\ \bibinfo {pages} {93} (\bibinfo {year} {1974})}\BibitemShut {NoStop}%
\bibitem [{\citenamefont {Abou-Chacra}\ \emph {et~al.}(1973)\citenamefont
  {Abou-Chacra}, \citenamefont {Thouless},\ and\ \citenamefont
  {Anderson}}]{r16}%
  \BibitemOpen
  \bibfield  {author} {\bibinfo {author} {\bibfnamefont {R.}~\bibnamefont
  {Abou-Chacra}}, \bibinfo {author} {\bibfnamefont {D.~J.}\ \bibnamefont
  {Thouless}}, \ and\ \bibinfo {author} {\bibfnamefont {P.~W.}\ \bibnamefont
  {Anderson}},\ }\href {\doibase 10.1088/0022-3719/6/10/009} {\bibfield
  {journal} {\bibinfo  {journal} {Journal of Physics C: Solid State Physics}\
  }\textbf {\bibinfo {volume} {6}},\ \bibinfo {pages} {009} (\bibinfo {year}
  {1973})}\BibitemShut {NoStop}%
\bibitem [{\citenamefont {Piao}\ and\ \citenamefont {Park}(2022)}]{web1}%
  \BibitemOpen
  \bibfield  {author} {\bibinfo {author} {\bibfnamefont {X.}~\bibnamefont
  {Piao}}\ and\ \bibinfo {author} {\bibfnamefont {N.}~\bibnamefont {Park}},\
  }\href {\doibase 10.1021/acsphotonics.2c00032} {\bibfield  {journal}
  {\bibinfo  {journal} {ACS Photonics}\ }\textbf {\bibinfo {volume} {9}},\
  \bibinfo {pages} {1655} (\bibinfo {year} {2022})}\BibitemShut {NoStop}%
\bibitem [{\citenamefont {Corona-Patricio}\ \emph {et~al.}(2019)\citenamefont
  {Corona-Patricio}, \citenamefont {Kuhl}, \citenamefont {Mortessagne},
  \citenamefont {Vignolo},\ and\ \citenamefont {Tessieri}}]{web2}%
  \BibitemOpen
  \bibfield  {author} {\bibinfo {author} {\bibfnamefont {G.}~\bibnamefont
  {Corona-Patricio}}, \bibinfo {author} {\bibfnamefont {U.}~\bibnamefont
  {Kuhl}}, \bibinfo {author} {\bibfnamefont {F.}~\bibnamefont {Mortessagne}},
  \bibinfo {author} {\bibfnamefont {P.}~\bibnamefont {Vignolo}}, \ and\
  \bibinfo {author} {\bibfnamefont {L.}~\bibnamefont {Tessieri}},\ }\href@noop
  {} {\bibfield  {journal} {\bibinfo  {journal} {New Journal Of Physics}\
  }\textbf {\bibinfo {volume} {21}} (\bibinfo {year} {2019})}\BibitemShut
  {NoStop}%
\bibitem [{\citenamefont {Sutradhar}\ \emph {et~al.}(2019)\citenamefont
  {Sutradhar}, \citenamefont {Mukerjee}, \citenamefont {Pandit},\ and\
  \citenamefont {Banerjee}}]{web3}%
  \BibitemOpen
  \bibfield  {author} {\bibinfo {author} {\bibfnamefont {J.}~\bibnamefont
  {Sutradhar}}, \bibinfo {author} {\bibfnamefont {S.}~\bibnamefont {Mukerjee}},
  \bibinfo {author} {\bibfnamefont {R.}~\bibnamefont {Pandit}}, \ and\ \bibinfo
  {author} {\bibfnamefont {S.}~\bibnamefont {Banerjee}},\ }\href {\doibase
  10.1103/PhysRevB.99.224204} {\bibfield  {journal} {\bibinfo  {journal}
  {Physical Review B}\ }\textbf {\bibinfo {volume} {99}} (\bibinfo {year}
  {2019}),\ 10.1103/PhysRevB.99.224204}\BibitemShut {NoStop}%
\bibitem [{\citenamefont {Agarwal}\ \emph {et~al.}(2017)\citenamefont
  {Agarwal}, \citenamefont {Ganeshan},\ and\ \citenamefont {Bhatt}}]{web4}%
  \BibitemOpen
  \bibfield  {author} {\bibinfo {author} {\bibfnamefont {K.}~\bibnamefont
  {Agarwal}}, \bibinfo {author} {\bibfnamefont {S.}~\bibnamefont {Ganeshan}}, \
  and\ \bibinfo {author} {\bibfnamefont {R.~N.}\ \bibnamefont {Bhatt}},\
  }\href@noop {} {\bibfield  {journal} {\bibinfo  {journal} {Physical Review
  B}\ }\textbf {\bibinfo {volume} {96}} (\bibinfo {year} {2017})}\BibitemShut
  {NoStop}%
\bibitem [{\citenamefont {Todorov}(2002)}]{web5}%
  \BibitemOpen
  \bibfield  {author} {\bibinfo {author} {\bibfnamefont {T.}~\bibnamefont
  {Todorov}},\ }\href {\doibase 10.1088/0953-8984/14/11/314} {\bibfield
  {journal} {\bibinfo  {journal} {Journal of Physics-Condensed Matter}\
  }\textbf {\bibinfo {volume} {14}},\ \bibinfo {pages} {3049} (\bibinfo {year}
  {2002})}\BibitemShut {NoStop}%
\bibitem [{\citenamefont {Furusaki}(1999)}]{web6}%
  \BibitemOpen
  \bibfield  {author} {\bibinfo {author} {\bibfnamefont {A.}~\bibnamefont
  {Furusaki}},\ }\href {\doibase 10.1103/PhysRevLett.82.604} {\bibfield
  {journal} {\bibinfo  {journal} {Physical Review Letters}\ }\textbf {\bibinfo
  {volume} {82}},\ \bibinfo {pages} {604} (\bibinfo {year} {1999})}\BibitemShut
  {NoStop}%
\bibitem [{\citenamefont {Sugiyama}\ and\ \citenamefont
  {Nagaosa}(1993)}]{web7}%
  \BibitemOpen
  \bibfield  {author} {\bibinfo {author} {\bibfnamefont {T.}~\bibnamefont
  {Sugiyama}}\ and\ \bibinfo {author} {\bibfnamefont {N.}~\bibnamefont
  {Nagaosa}},\ }\href {\doibase 10.1103/PhysRevLett.70.1980} {\bibfield
  {journal} {\bibinfo  {journal} {Physical Review Letters}\ }\textbf {\bibinfo
  {volume} {70}},\ \bibinfo {pages} {1980} (\bibinfo {year}
  {1993})}\BibitemShut {NoStop}%
\bibitem [{\citenamefont {Chabanov}\ \emph {et~al.}(2000)\citenamefont
  {Chabanov}, \citenamefont {Stoytchev},\ and\ \citenamefont {Genack}}]{r2}%
  \BibitemOpen
  \bibfield  {author} {\bibinfo {author} {\bibfnamefont {A.~A.}\ \bibnamefont
  {Chabanov}}, \bibinfo {author} {\bibfnamefont {M.}~\bibnamefont {Stoytchev}},
  \ and\ \bibinfo {author} {\bibfnamefont {A.~Z.}\ \bibnamefont {Genack}},\
  }\href {\doibase 10.1038/35009055} {\bibfield  {journal} {\bibinfo  {journal}
  {Nature}\ }\textbf {\bibinfo {volume} {404}},\ \bibinfo {pages} {850}
  (\bibinfo {year} {2000})}\BibitemShut {NoStop}%
\bibitem [{\citenamefont {Weaver}(1990)}]{r3}%
  \BibitemOpen
  \bibfield  {author} {\bibinfo {author} {\bibfnamefont {R.}~\bibnamefont
  {Weaver}},\ }\href {\doibase 10.1016/0165-2125(90)90034-2} {\bibfield
  {journal} {\bibinfo  {journal} {Wave Motion}\ }\textbf {\bibinfo {volume}
  {12}},\ \bibinfo {pages} {129} (\bibinfo {year} {1990})}\BibitemShut
  {NoStop}%
\bibitem [{\citenamefont {Hu}\ \emph {et~al.}(2008)\citenamefont {Hu},
  \citenamefont {Strybulevych}, \citenamefont {Page}, \citenamefont
  {Skipetrov},\ and\ \citenamefont {van Tiggelen}}]{r4}%
  \BibitemOpen
  \bibfield  {author} {\bibinfo {author} {\bibfnamefont {H.}~\bibnamefont
  {Hu}}, \bibinfo {author} {\bibfnamefont {A.}~\bibnamefont {Strybulevych}},
  \bibinfo {author} {\bibfnamefont {J.~H.}\ \bibnamefont {Page}}, \bibinfo
  {author} {\bibfnamefont {S.~E.}\ \bibnamefont {Skipetrov}}, \ and\ \bibinfo
  {author} {\bibfnamefont {B.~A.}\ \bibnamefont {van Tiggelen}},\ }\href
  {\doibase 10.1038/nphys1101} {\bibfield  {journal} {\bibinfo  {journal}
  {Nature Physics}\ }\textbf {\bibinfo {volume} {4}},\ \bibinfo {pages} {945}
  (\bibinfo {year} {2008})}\BibitemShut {NoStop}%
\bibitem [{\citenamefont {Schwartz}\ \emph {et~al.}(2007)\citenamefont
  {Schwartz}, \citenamefont {Bartal}, \citenamefont {Fishman},\ and\
  \citenamefont {Segev}}]{r5}%
  \BibitemOpen
  \bibfield  {author} {\bibinfo {author} {\bibfnamefont {T.}~\bibnamefont
  {Schwartz}}, \bibinfo {author} {\bibfnamefont {G.}~\bibnamefont {Bartal}},
  \bibinfo {author} {\bibfnamefont {S.}~\bibnamefont {Fishman}}, \ and\
  \bibinfo {author} {\bibfnamefont {M.}~\bibnamefont {Segev}},\ }\href
  {\doibase 10.1038/nature05623} {\bibfield  {journal} {\bibinfo  {journal}
  {Nature}\ }\textbf {\bibinfo {volume} {446}},\ \bibinfo {pages} {52}
  (\bibinfo {year} {2007})}\BibitemShut {NoStop}%
\bibitem [{\citenamefont {Chabé}\ \emph {et~al.}(2008)\citenamefont {Chabé},
  \citenamefont {Lemarié}, \citenamefont {Grémaud}, \citenamefont {Delande},
  \citenamefont {Szriftgiser},\ and\ \citenamefont {Garreau}}]{r6}%
  \BibitemOpen
  \bibfield  {author} {\bibinfo {author} {\bibfnamefont {J.}~\bibnamefont
  {Chabé}}, \bibinfo {author} {\bibfnamefont {G.}~\bibnamefont {Lemarié}},
  \bibinfo {author} {\bibfnamefont {B.}~\bibnamefont {Grémaud}}, \bibinfo
  {author} {\bibfnamefont {D.}~\bibnamefont {Delande}}, \bibinfo {author}
  {\bibfnamefont {P.}~\bibnamefont {Szriftgiser}}, \ and\ \bibinfo {author}
  {\bibfnamefont {J.~C.}\ \bibnamefont {Garreau}},\ }\href {\doibase
  10.1103/PhysRevLett.101.255702} {\bibfield  {journal} {\bibinfo  {journal}
  {Physical Review Letters}\ }\textbf {\bibinfo {volume} {101}},\ \bibinfo
  {pages} {255702} (\bibinfo {year} {2008})}\BibitemShut {NoStop}%
\bibitem [{\citenamefont {Lagendijk}\ \emph {et~al.}(2009)\citenamefont
  {Lagendijk}, \citenamefont {van Tiggelen},\ and\ \citenamefont
  {Wiersma}}]{r7}%
  \BibitemOpen
  \bibfield  {author} {\bibinfo {author} {\bibfnamefont {A.}~\bibnamefont
  {Lagendijk}}, \bibinfo {author} {\bibfnamefont {B.}~\bibnamefont {van
  Tiggelen}}, \ and\ \bibinfo {author} {\bibfnamefont {D.~S.}\ \bibnamefont
  {Wiersma}},\ }\href {\doibase 10.1063/1.3206091} {\bibfield  {journal}
  {\bibinfo  {journal} {Physics Today}\ }\textbf {\bibinfo {volume} {62}},\
  \bibinfo {pages} {24} (\bibinfo {year} {2009})}\BibitemShut {NoStop}%
\bibitem [{\citenamefont {Li}\ \emph {et~al.}(2009)\citenamefont {Li},
  \citenamefont {Chu}, \citenamefont {Jain},\ and\ \citenamefont
  {Shen}}]{li_topological_2009}%
  \BibitemOpen
  \bibfield  {author} {\bibinfo {author} {\bibfnamefont {J.}~\bibnamefont
  {Li}}, \bibinfo {author} {\bibfnamefont {R.-L.}\ \bibnamefont {Chu}},
  \bibinfo {author} {\bibfnamefont {J.~K.}\ \bibnamefont {Jain}}, \ and\
  \bibinfo {author} {\bibfnamefont {S.-Q.}\ \bibnamefont {Shen}},\ }\href
  {\doibase 10.1103/PhysRevLett.102.136806} {\bibfield  {journal} {\bibinfo
  {journal} {Physical Review Letters}\ }\textbf {\bibinfo {volume} {102}},\
  \bibinfo {pages} {136806} (\bibinfo {year} {2009})}\BibitemShut {NoStop}%
\bibitem [{\citenamefont {Groth}\ \emph {et~al.}(2009)\citenamefont {Groth},
  \citenamefont {Wimmer}, \citenamefont {Akhmerov}, \citenamefont
  {Tworzyd\l{}o},\ and\ \citenamefont {Beenakker}}]{groth_theory_2009}%
  \BibitemOpen
  \bibfield  {author} {\bibinfo {author} {\bibfnamefont {C.~W.}\ \bibnamefont
  {Groth}}, \bibinfo {author} {\bibfnamefont {M.}~\bibnamefont {Wimmer}},
  \bibinfo {author} {\bibfnamefont {A.~R.}\ \bibnamefont {Akhmerov}}, \bibinfo
  {author} {\bibfnamefont {J.}~\bibnamefont {Tworzyd\l{}o}}, \ and\ \bibinfo
  {author} {\bibfnamefont {C.~W.~J.}\ \bibnamefont {Beenakker}},\ }\href
  {\doibase 10.1103/PhysRevLett.103.196805} {\bibfield  {journal} {\bibinfo
  {journal} {Phys. Rev. Lett.}\ }\textbf {\bibinfo {volume} {103}},\ \bibinfo
  {pages} {196805} (\bibinfo {year} {2009})}\BibitemShut {NoStop}%
\bibitem [{\citenamefont {Song}\ \emph {et~al.}(2012)\citenamefont {Song},
  \citenamefont {Liu}, \citenamefont {Jiang}, \citenamefont {Sun},\ and\
  \citenamefont {Xie}}]{song_dependence_2012}%
  \BibitemOpen
  \bibfield  {author} {\bibinfo {author} {\bibfnamefont {J.}~\bibnamefont
  {Song}}, \bibinfo {author} {\bibfnamefont {H.}~\bibnamefont {Liu}}, \bibinfo
  {author} {\bibfnamefont {H.}~\bibnamefont {Jiang}}, \bibinfo {author}
  {\bibfnamefont {Q.-f.}\ \bibnamefont {Sun}}, \ and\ \bibinfo {author}
  {\bibfnamefont {X.~C.}\ \bibnamefont {Xie}},\ }\href {\doibase
  10.1103/PhysRevB.85.195125} {\bibfield  {journal} {\bibinfo  {journal} {Phys.
  Rev. B}\ }\textbf {\bibinfo {volume} {85}},\ \bibinfo {pages} {195125}
  (\bibinfo {year} {2012})}\BibitemShut {NoStop}%
\bibitem [{\citenamefont {Kotakoski}\ \emph {et~al.}(2011)\citenamefont
  {Kotakoski}, \citenamefont {Krasheninnikov}, \citenamefont {Kaiser},\ and\
  \citenamefont {Meyer}}]{kotakoski_from_2011}%
  \BibitemOpen
  \bibfield  {author} {\bibinfo {author} {\bibfnamefont {J.}~\bibnamefont
  {Kotakoski}}, \bibinfo {author} {\bibfnamefont {A.~V.}\ \bibnamefont
  {Krasheninnikov}}, \bibinfo {author} {\bibfnamefont {U.}~\bibnamefont
  {Kaiser}}, \ and\ \bibinfo {author} {\bibfnamefont {J.~C.}\ \bibnamefont
  {Meyer}},\ }\href {\doibase 10.1103/PhysRevLett.106.105505} {\bibfield
  {journal} {\bibinfo  {journal} {Phys. Rev. Lett.}\ }\textbf {\bibinfo
  {volume} {106}},\ \bibinfo {pages} {105505} (\bibinfo {year}
  {2011})}\BibitemShut {NoStop}%
\bibitem [{\citenamefont {Eder}\ \emph {et~al.}(2014)\citenamefont {Eder},
  \citenamefont {Kotakoski}, \citenamefont {Kaiser},\ and\ \citenamefont
  {Meyer}}]{eder_journey_2014}%
  \BibitemOpen
  \bibfield  {author} {\bibinfo {author} {\bibfnamefont {F.~R.}\ \bibnamefont
  {Eder}}, \bibinfo {author} {\bibfnamefont {J.}~\bibnamefont {Kotakoski}},
  \bibinfo {author} {\bibfnamefont {U.}~\bibnamefont {Kaiser}}, \ and\ \bibinfo
  {author} {\bibfnamefont {J.~C.}\ \bibnamefont {Meyer}},\ }\href {\doibase
  10.1038/srep04060} {\bibfield  {journal} {\bibinfo  {journal} {Scientific
  Reports}\ }\textbf {\bibinfo {volume} {4}},\ \bibinfo {pages} {4060}
  (\bibinfo {year} {2014})}\BibitemShut {NoStop}%
\bibitem [{\citenamefont {Joo}\ \emph {et~al.}(2017)\citenamefont {Joo},
  \citenamefont {Lee}, \citenamefont {Jang}, \citenamefont {Kang},
  \citenamefont {Kwon}, \citenamefont {Chung}, \citenamefont {Lee},
  \citenamefont {Kim}, \citenamefont {Kim}, \citenamefont {Yang} \emph
  {et~al.}}]{joo2017realization}%
  \BibitemOpen
  \bibfield  {author} {\bibinfo {author} {\bibfnamefont {W.-J.}\ \bibnamefont
  {Joo}}, \bibinfo {author} {\bibfnamefont {J.-H.}\ \bibnamefont {Lee}},
  \bibinfo {author} {\bibfnamefont {Y.}~\bibnamefont {Jang}}, \bibinfo {author}
  {\bibfnamefont {S.-G.}\ \bibnamefont {Kang}}, \bibinfo {author}
  {\bibfnamefont {Y.-N.}\ \bibnamefont {Kwon}}, \bibinfo {author}
  {\bibfnamefont {J.}~\bibnamefont {Chung}}, \bibinfo {author} {\bibfnamefont
  {S.}~\bibnamefont {Lee}}, \bibinfo {author} {\bibfnamefont {C.}~\bibnamefont
  {Kim}}, \bibinfo {author} {\bibfnamefont {T.-H.}\ \bibnamefont {Kim}},
  \bibinfo {author} {\bibfnamefont {C.-W.}\ \bibnamefont {Yang}},  \emph
  {et~al.},\ }\href@noop {} {\bibfield  {journal} {\bibinfo  {journal} {Science
  advances}\ }\textbf {\bibinfo {volume} {3}},\ \bibinfo {pages} {e1601821}
  (\bibinfo {year} {2017})}\BibitemShut {NoStop}%
\bibitem [{\citenamefont {Toh}\ \emph {et~al.}(2020)\citenamefont {Toh},
  \citenamefont {Zhang}, \citenamefont {Lin}, \citenamefont {Mayorov},
  \citenamefont {Wang}, \citenamefont {Orofeo}, \citenamefont {Ferry},
  \citenamefont {Andersen}, \citenamefont {Kakenov}, \citenamefont {Guo},
  \citenamefont {Abidi}, \citenamefont {Sims}, \citenamefont {Suenaga},
  \citenamefont {Pantelides},\ and\ \citenamefont
  {Özyilmaz}}]{toh_synthesis_2020}%
  \BibitemOpen
  \bibfield  {author} {\bibinfo {author} {\bibfnamefont {C.-T.}\ \bibnamefont
  {Toh}}, \bibinfo {author} {\bibfnamefont {H.}~\bibnamefont {Zhang}}, \bibinfo
  {author} {\bibfnamefont {J.}~\bibnamefont {Lin}}, \bibinfo {author}
  {\bibfnamefont {A.~S.}\ \bibnamefont {Mayorov}}, \bibinfo {author}
  {\bibfnamefont {Y.-P.}\ \bibnamefont {Wang}}, \bibinfo {author}
  {\bibfnamefont {C.~M.}\ \bibnamefont {Orofeo}}, \bibinfo {author}
  {\bibfnamefont {D.~B.}\ \bibnamefont {Ferry}}, \bibinfo {author}
  {\bibfnamefont {H.}~\bibnamefont {Andersen}}, \bibinfo {author}
  {\bibfnamefont {N.}~\bibnamefont {Kakenov}}, \bibinfo {author} {\bibfnamefont
  {Z.}~\bibnamefont {Guo}}, \bibinfo {author} {\bibfnamefont {I.~H.}\
  \bibnamefont {Abidi}}, \bibinfo {author} {\bibfnamefont {H.}~\bibnamefont
  {Sims}}, \bibinfo {author} {\bibfnamefont {K.}~\bibnamefont {Suenaga}},
  \bibinfo {author} {\bibfnamefont {S.~T.}\ \bibnamefont {Pantelides}}, \ and\
  \bibinfo {author} {\bibfnamefont {B.}~\bibnamefont {Özyilmaz}},\ }\href
  {\doibase 10.1038/s41586-019-1871-2} {\bibfield  {journal} {\bibinfo
  {journal} {Nature}\ }\textbf {\bibinfo {volume} {577}},\ \bibinfo {pages}
  {199} (\bibinfo {year} {2020})}\BibitemShut {NoStop}%
\bibitem [{\citenamefont {Tian}\ \emph {et~al.}(2023)\citenamefont {Tian},
  \citenamefont {Ma}, \citenamefont {Li}, \citenamefont {Cheng}, \citenamefont
  {Ning}, \citenamefont {Han}, \citenamefont {Xu}, \citenamefont {Zhang},
  \citenamefont {Zhao}, \citenamefont {Li}, \citenamefont {Zou}, \citenamefont
  {Liao}, \citenamefont {Yu}, \citenamefont {Li}, \citenamefont {Wang},
  \citenamefont {Liu}, \citenamefont {Li}, \citenamefont {Huang}, \citenamefont
  {Yao}, \citenamefont {Ding}, \citenamefont {Guo}, \citenamefont {Huang},
  \citenamefont {Lu}, \citenamefont {Han}, \citenamefont {Wang}, \citenamefont
  {Cheng}, \citenamefont {Liu}, \citenamefont {Xu}, \citenamefont {Liu},
  \citenamefont {Gao}, \citenamefont {Jiang}, \citenamefont {Lin},
  \citenamefont {Zhao}, \citenamefont {Wang}, \citenamefont {Bai},
  \citenamefont {Fu}, \citenamefont {Wang}, \citenamefont {Li}, \citenamefont
  {Lei}, \citenamefont {Zhang}, \citenamefont {Hou}, \citenamefont {Pei},
  \citenamefont {Pennycook}, \citenamefont {Wang}, \citenamefont {Chen},
  \citenamefont {Zhou},\ and\ \citenamefont {Liu}}]{tian_disorder-tuned_2023}%
  \BibitemOpen
  \bibfield  {author} {\bibinfo {author} {\bibfnamefont {H.}~\bibnamefont
  {Tian}}, \bibinfo {author} {\bibfnamefont {Y.}~\bibnamefont {Ma}}, \bibinfo
  {author} {\bibfnamefont {Z.}~\bibnamefont {Li}}, \bibinfo {author}
  {\bibfnamefont {M.}~\bibnamefont {Cheng}}, \bibinfo {author} {\bibfnamefont
  {S.}~\bibnamefont {Ning}}, \bibinfo {author} {\bibfnamefont {E.}~\bibnamefont
  {Han}}, \bibinfo {author} {\bibfnamefont {M.}~\bibnamefont {Xu}}, \bibinfo
  {author} {\bibfnamefont {P.-F.}\ \bibnamefont {Zhang}}, \bibinfo {author}
  {\bibfnamefont {K.}~\bibnamefont {Zhao}}, \bibinfo {author} {\bibfnamefont
  {R.}~\bibnamefont {Li}}, \bibinfo {author} {\bibfnamefont {Y.}~\bibnamefont
  {Zou}}, \bibinfo {author} {\bibfnamefont {P.}~\bibnamefont {Liao}}, \bibinfo
  {author} {\bibfnamefont {S.}~\bibnamefont {Yu}}, \bibinfo {author}
  {\bibfnamefont {X.}~\bibnamefont {Li}}, \bibinfo {author} {\bibfnamefont
  {J.}~\bibnamefont {Wang}}, \bibinfo {author} {\bibfnamefont {S.}~\bibnamefont
  {Liu}}, \bibinfo {author} {\bibfnamefont {Y.}~\bibnamefont {Li}}, \bibinfo
  {author} {\bibfnamefont {X.}~\bibnamefont {Huang}}, \bibinfo {author}
  {\bibfnamefont {Z.}~\bibnamefont {Yao}}, \bibinfo {author} {\bibfnamefont
  {D.}~\bibnamefont {Ding}}, \bibinfo {author} {\bibfnamefont {J.}~\bibnamefont
  {Guo}}, \bibinfo {author} {\bibfnamefont {Y.}~\bibnamefont {Huang}}, \bibinfo
  {author} {\bibfnamefont {J.}~\bibnamefont {Lu}}, \bibinfo {author}
  {\bibfnamefont {Y.}~\bibnamefont {Han}}, \bibinfo {author} {\bibfnamefont
  {Z.}~\bibnamefont {Wang}}, \bibinfo {author} {\bibfnamefont {Z.~G.}\
  \bibnamefont {Cheng}}, \bibinfo {author} {\bibfnamefont {J.}~\bibnamefont
  {Liu}}, \bibinfo {author} {\bibfnamefont {Z.}~\bibnamefont {Xu}}, \bibinfo
  {author} {\bibfnamefont {K.}~\bibnamefont {Liu}}, \bibinfo {author}
  {\bibfnamefont {P.}~\bibnamefont {Gao}}, \bibinfo {author} {\bibfnamefont
  {Y.}~\bibnamefont {Jiang}}, \bibinfo {author} {\bibfnamefont
  {L.}~\bibnamefont {Lin}}, \bibinfo {author} {\bibfnamefont {X.}~\bibnamefont
  {Zhao}}, \bibinfo {author} {\bibfnamefont {L.}~\bibnamefont {Wang}}, \bibinfo
  {author} {\bibfnamefont {X.}~\bibnamefont {Bai}}, \bibinfo {author}
  {\bibfnamefont {W.}~\bibnamefont {Fu}}, \bibinfo {author} {\bibfnamefont
  {J.-Y.}\ \bibnamefont {Wang}}, \bibinfo {author} {\bibfnamefont
  {M.}~\bibnamefont {Li}}, \bibinfo {author} {\bibfnamefont {T.}~\bibnamefont
  {Lei}}, \bibinfo {author} {\bibfnamefont {Y.}~\bibnamefont {Zhang}}, \bibinfo
  {author} {\bibfnamefont {Y.}~\bibnamefont {Hou}}, \bibinfo {author}
  {\bibfnamefont {J.}~\bibnamefont {Pei}}, \bibinfo {author} {\bibfnamefont
  {S.~J.}\ \bibnamefont {Pennycook}}, \bibinfo {author} {\bibfnamefont
  {E.}~\bibnamefont {Wang}}, \bibinfo {author} {\bibfnamefont {J.}~\bibnamefont
  {Chen}}, \bibinfo {author} {\bibfnamefont {W.}~\bibnamefont {Zhou}}, \ and\
  \bibinfo {author} {\bibfnamefont {L.}~\bibnamefont {Liu}},\ }\href {\doibase
  10.1038/s41586-022-05617-w} {\bibfield  {journal} {\bibinfo  {journal}
  {Nature}\ }\textbf {\bibinfo {volume} {615}},\ \bibinfo {pages} {56}
  (\bibinfo {year} {2023})}\BibitemShut {NoStop}%
\bibitem [{\citenamefont {Zhao}\ \emph {et~al.}(2019)\citenamefont {Zhao},
  \citenamefont {Chen}, \citenamefont {Wang}, \citenamefont {Qiu},\ and\
  \citenamefont {Guo}}]{Zhao_two_2019}%
  \BibitemOpen
  \bibfield  {author} {\bibinfo {author} {\bibfnamefont {H.}~\bibnamefont
  {Zhao}}, \bibinfo {author} {\bibfnamefont {X.}~\bibnamefont {Chen}}, \bibinfo
  {author} {\bibfnamefont {G.}~\bibnamefont {Wang}}, \bibinfo {author}
  {\bibfnamefont {Y.}~\bibnamefont {Qiu}}, \ and\ \bibinfo {author}
  {\bibfnamefont {L.}~\bibnamefont {Guo}},\ }\href {\doibase
  10.1088/2053-1583/ab1169} {\bibfield  {journal} {\bibinfo  {journal} {2D
  Materials}\ }\textbf {\bibinfo {volume} {6}},\ \bibinfo {pages} {032002}
  (\bibinfo {year} {2019})}\BibitemShut {NoStop}%
\bibitem [{\citenamefont {Hong}\ \emph {et~al.}(2020)\citenamefont {Hong},
  \citenamefont {Lee}, \citenamefont {Lee}, \citenamefont {Lee}, \citenamefont
  {Ma}, \citenamefont {Kim}, \citenamefont {Yoon}, \citenamefont {Ihm},
  \citenamefont {Kim}, \citenamefont {Shin} \emph {et~al.}}]{hong2020ultralow}%
  \BibitemOpen
  \bibfield  {author} {\bibinfo {author} {\bibfnamefont {S.}~\bibnamefont
  {Hong}}, \bibinfo {author} {\bibfnamefont {C.-S.}\ \bibnamefont {Lee}},
  \bibinfo {author} {\bibfnamefont {M.-H.}\ \bibnamefont {Lee}}, \bibinfo
  {author} {\bibfnamefont {Y.}~\bibnamefont {Lee}}, \bibinfo {author}
  {\bibfnamefont {K.~Y.}\ \bibnamefont {Ma}}, \bibinfo {author} {\bibfnamefont
  {G.}~\bibnamefont {Kim}}, \bibinfo {author} {\bibfnamefont {S.~I.}\
  \bibnamefont {Yoon}}, \bibinfo {author} {\bibfnamefont {K.}~\bibnamefont
  {Ihm}}, \bibinfo {author} {\bibfnamefont {K.-J.}\ \bibnamefont {Kim}},
  \bibinfo {author} {\bibfnamefont {T.~J.}\ \bibnamefont {Shin}},  \emph
  {et~al.},\ }\href@noop {} {\bibfield  {journal} {\bibinfo  {journal}
  {Nature}\ }\textbf {\bibinfo {volume} {582}},\ \bibinfo {pages} {511}
  (\bibinfo {year} {2020})}\BibitemShut {NoStop}%
\bibitem [{\citenamefont {Shin}\ \emph {et~al.}(2021)\citenamefont {Shin},
  \citenamefont {Wang}, \citenamefont {Han}, \citenamefont {Lin}, \citenamefont
  {O'Hara}, \citenamefont {Chen}, \citenamefont {Lin},\ and\ \citenamefont
  {Pantelides}}]{shin_preferential_2021}%
  \BibitemOpen
  \bibfield  {author} {\bibinfo {author} {\bibfnamefont {D.}~\bibnamefont
  {Shin}}, \bibinfo {author} {\bibfnamefont {G.}~\bibnamefont {Wang}}, \bibinfo
  {author} {\bibfnamefont {M.}~\bibnamefont {Han}}, \bibinfo {author}
  {\bibfnamefont {Z.}~\bibnamefont {Lin}}, \bibinfo {author} {\bibfnamefont
  {A.}~\bibnamefont {O'Hara}}, \bibinfo {author} {\bibfnamefont
  {F.}~\bibnamefont {Chen}}, \bibinfo {author} {\bibfnamefont {J.}~\bibnamefont
  {Lin}}, \ and\ \bibinfo {author} {\bibfnamefont {S.~T.}\ \bibnamefont
  {Pantelides}},\ }\href {\doibase 10.1103/PhysRevMaterials.5.044002}
  {\bibfield  {journal} {\bibinfo  {journal} {Phys. Rev. Mater.}\ }\textbf
  {\bibinfo {volume} {5}},\ \bibinfo {pages} {044002} (\bibinfo {year}
  {2021})}\BibitemShut {NoStop}%
\bibitem [{\citenamefont {Zhang}\ \emph
  {et~al.}(2022{\natexlab{a}})\citenamefont {Zhang}, \citenamefont {Wang},
  \citenamefont {Zhang}, \citenamefont {Zhang}, \citenamefont {Du},\ and\
  \citenamefont {Pantelides}}]{zhang_structure_2022}%
  \BibitemOpen
  \bibfield  {author} {\bibinfo {author} {\bibfnamefont {Y.-T.}\ \bibnamefont
  {Zhang}}, \bibinfo {author} {\bibfnamefont {Y.-P.}\ \bibnamefont {Wang}},
  \bibinfo {author} {\bibfnamefont {X.}~\bibnamefont {Zhang}}, \bibinfo
  {author} {\bibfnamefont {Y.-Y.}\ \bibnamefont {Zhang}}, \bibinfo {author}
  {\bibfnamefont {S.}~\bibnamefont {Du}}, \ and\ \bibinfo {author}
  {\bibfnamefont {S.~T.}\ \bibnamefont {Pantelides}},\ }\href {\doibase
  10.1021/acs.nanolett.2c02542} {\bibfield  {journal} {\bibinfo  {journal}
  {Nano Letters}\ }\textbf {\bibinfo {volume} {22}},\ \bibinfo {pages} {8018}
  (\bibinfo {year} {2022}{\natexlab{a}})}\BibitemShut {NoStop}%
\bibitem [{\citenamefont {Kubo}(1957)}]{Kubo1957}%
  \BibitemOpen
  \bibfield  {author} {\bibinfo {author} {\bibfnamefont {R.}~\bibnamefont
  {Kubo}},\ }\href {\doibase 10.1143/JPSJ.12.570} {\bibfield  {journal}
  {\bibinfo  {journal} {Journal of the Physical Society of Japan}\ }\textbf
  {\bibinfo {volume} {12}},\ \bibinfo {pages} {570} (\bibinfo {year}
  {1957})}\BibitemShut {NoStop}%
\bibitem [{\citenamefont {Greenwood}(1958)}]{Greenwood1958}%
  \BibitemOpen
  \bibfield  {author} {\bibinfo {author} {\bibfnamefont {D.~A.}\ \bibnamefont
  {Greenwood}},\ }\href {\doibase 10.1088/0370-1328/71/4/306} {\bibfield
  {journal} {\bibinfo  {journal} {Proceedings of the Physical Society}\
  }\textbf {\bibinfo {volume} {71}},\ \bibinfo {pages} {585} (\bibinfo {year}
  {1958})}\BibitemShut {NoStop}%
\bibitem [{\citenamefont {Kapko}\ \emph {et~al.}(2010)\citenamefont {Kapko},
  \citenamefont {Drabold},\ and\ \citenamefont
  {Thorpe}}]{kapko_electronic_2010}%
  \BibitemOpen
  \bibfield  {author} {\bibinfo {author} {\bibfnamefont {V.}~\bibnamefont
  {Kapko}}, \bibinfo {author} {\bibfnamefont {D.~A.}\ \bibnamefont {Drabold}},
  \ and\ \bibinfo {author} {\bibfnamefont {M.~F.}\ \bibnamefont {Thorpe}},\
  }\href {\doibase 10.1002/pssb.200945581} {\bibfield  {journal} {\bibinfo
  {journal} {physica status solidi (b)}\ }\textbf {\bibinfo {volume} {247}},\
  \bibinfo {pages} {1197} (\bibinfo {year} {2010})}\BibitemShut {NoStop}%
\bibitem [{\citenamefont {Van~Tuan}\ \emph {et~al.}(2012)\citenamefont
  {Van~Tuan}, \citenamefont {Kumar}, \citenamefont {Roche}, \citenamefont
  {Ortmann}, \citenamefont {Thorpe},\ and\ \citenamefont
  {Ordejon}}]{van_tuan_insulating_2012}%
  \BibitemOpen
  \bibfield  {author} {\bibinfo {author} {\bibfnamefont {D.}~\bibnamefont
  {Van~Tuan}}, \bibinfo {author} {\bibfnamefont {A.}~\bibnamefont {Kumar}},
  \bibinfo {author} {\bibfnamefont {S.}~\bibnamefont {Roche}}, \bibinfo
  {author} {\bibfnamefont {F.}~\bibnamefont {Ortmann}}, \bibinfo {author}
  {\bibfnamefont {M.~F.}\ \bibnamefont {Thorpe}}, \ and\ \bibinfo {author}
  {\bibfnamefont {P.}~\bibnamefont {Ordejon}},\ }\href {\doibase
  10.1103/PhysRevB.86.121408} {\bibfield  {journal} {\bibinfo  {journal}
  {Physical Review B}\ }\textbf {\bibinfo {volume} {86}},\ \bibinfo {pages}
  {121408} (\bibinfo {year} {2012})}\BibitemShut {NoStop}%
\bibitem [{\citenamefont {Thapa}\ \emph {et~al.}(2022)\citenamefont {Thapa},
  \citenamefont {Ugwumadu}, \citenamefont {Nepal}, \citenamefont {Trembly},\
  and\ \citenamefont {Drabold}}]{thapa_ab_2022}%
  \BibitemOpen
  \bibfield  {author} {\bibinfo {author} {\bibfnamefont {R.}~\bibnamefont
  {Thapa}}, \bibinfo {author} {\bibfnamefont {C.}~\bibnamefont {Ugwumadu}},
  \bibinfo {author} {\bibfnamefont {K.}~\bibnamefont {Nepal}}, \bibinfo
  {author} {\bibfnamefont {J.}~\bibnamefont {Trembly}}, \ and\ \bibinfo
  {author} {\bibfnamefont {D.}~\bibnamefont {Drabold}},\ }\href {\doibase
  10.1103/PhysRevLett.128.236402} {\bibfield  {journal} {\bibinfo  {journal}
  {Physical Review Letters}\ }\textbf {\bibinfo {volume} {128}},\ \bibinfo
  {pages} {236402} (\bibinfo {year} {2022})}\BibitemShut {NoStop}%
\bibitem [{Note1()}]{Note1}%
  \BibitemOpen
  \bibinfo {note} {Our model has a constraint that $\lambda <0.5$. This is
  because as $\lambda >0.5$, the wavefunction starts to delocalize as $\lambda
  $ continues to rise. This can be easily shown at $\lambda =1$, where the
  Hamiltonian is simply $ \protect \hat {H}(\lambda =1)=E_m\protect \hat
  {I}+\protect \hat {H}(\lambda =0) $ resulting in the same eigenstates as
  $\protect \hat {H}(\lambda =0)$, which returns to the $\lambda =0$
  case.}\BibitemShut {Stop}%
\bibitem [{\citenamefont {Li}\ and\ \citenamefont {Sarma}(2020)}]{r17}%
  \BibitemOpen
  \bibfield  {author} {\bibinfo {author} {\bibfnamefont {X.}~\bibnamefont
  {Li}}\ and\ \bibinfo {author} {\bibfnamefont {S.~D.}\ \bibnamefont {Sarma}},\
  }\href {\doibase 10.1103/PhysRevB.101.064203} {\bibfield  {journal} {\bibinfo
   {journal} {Physical Review B}\ }\textbf {\bibinfo {volume} {101}},\ \bibinfo
  {pages} {064203} (\bibinfo {year} {2020})}\BibitemShut {NoStop}%
\bibitem [{\citenamefont {Roy}\ \emph {et~al.}(2021)\citenamefont {Roy},
  \citenamefont {Mishra}, \citenamefont {Tanatar},\ and\ \citenamefont
  {Basu}}]{Roy2021}%
  \BibitemOpen
  \bibfield  {author} {\bibinfo {author} {\bibfnamefont {S.}~\bibnamefont
  {Roy}}, \bibinfo {author} {\bibfnamefont {T.}~\bibnamefont {Mishra}},
  \bibinfo {author} {\bibfnamefont {B.}~\bibnamefont {Tanatar}}, \ and\
  \bibinfo {author} {\bibfnamefont {S.}~\bibnamefont {Basu}},\ }\href {\doibase
  10.1103/PhysRevLett.126.106803} {\bibfield  {journal} {\bibinfo  {journal}
  {Physical Review Letters}\ }\textbf {\bibinfo {volume} {126}},\ \bibinfo
  {pages} {106803} (\bibinfo {year} {2021})}\BibitemShut {NoStop}%
\bibitem [{\citenamefont {Mott}\ and\ \citenamefont
  {Davis}(2012)}]{mott2012electronic}%
  \BibitemOpen
  \bibfield  {author} {\bibinfo {author} {\bibfnamefont {N.~F.}\ \bibnamefont
  {Mott}}\ and\ \bibinfo {author} {\bibfnamefont {E.~A.}\ \bibnamefont
  {Davis}},\ }\href@noop {} {\emph {\bibinfo {title} {Electronic processes in
  non-crystalline materials}}}\ (\bibinfo  {publisher} {Oxford university
  press},\ \bibinfo {year} {2012})\BibitemShut {NoStop}%
\bibitem [{\citenamefont {Carva}\ \emph {et~al.}(2010)\citenamefont {Carva},
  \citenamefont {Sanyal}, \citenamefont {Fransson},\ and\ \citenamefont
  {Eriksson}}]{carva2010defect}%
  \BibitemOpen
  \bibfield  {author} {\bibinfo {author} {\bibfnamefont {K.}~\bibnamefont
  {Carva}}, \bibinfo {author} {\bibfnamefont {B.}~\bibnamefont {Sanyal}},
  \bibinfo {author} {\bibfnamefont {J.}~\bibnamefont {Fransson}}, \ and\
  \bibinfo {author} {\bibfnamefont {O.}~\bibnamefont {Eriksson}},\ }\href@noop
  {} {\bibfield  {journal} {\bibinfo  {journal} {Physical Review B}\ }\textbf
  {\bibinfo {volume} {81}},\ \bibinfo {pages} {245405} (\bibinfo {year}
  {2010})}\BibitemShut {NoStop}%
\bibitem [{\citenamefont {Holmström}\ \emph {et~al.}(2011)\citenamefont
  {Holmström}, \citenamefont {Fransson}, \citenamefont {Eriksson},
  \citenamefont {Lizárraga}, \citenamefont {Sanyal}, \citenamefont
  {Bhandary},\ and\ \citenamefont {Katsnelson}}]{anomaly_carbon}%
  \BibitemOpen
  \bibfield  {author} {\bibinfo {author} {\bibfnamefont {E.}~\bibnamefont
  {Holmström}}, \bibinfo {author} {\bibfnamefont {J.}~\bibnamefont
  {Fransson}}, \bibinfo {author} {\bibfnamefont {O.}~\bibnamefont {Eriksson}},
  \bibinfo {author} {\bibfnamefont {R.}~\bibnamefont {Lizárraga}}, \bibinfo
  {author} {\bibfnamefont {B.}~\bibnamefont {Sanyal}}, \bibinfo {author}
  {\bibfnamefont {S.}~\bibnamefont {Bhandary}}, \ and\ \bibinfo {author}
  {\bibfnamefont {M.~I.}\ \bibnamefont {Katsnelson}},\ }\href {\doibase
  10.1103/PhysRevB.84.205414} {\bibfield  {journal} {\bibinfo  {journal}
  {Physical Review B}\ }\textbf {\bibinfo {volume} {84}},\ \bibinfo {pages}
  {205414} (\bibinfo {year} {2011})}\BibitemShut {NoStop}%
\bibitem [{\citenamefont {Khveshchenko}(2006)}]{khveshchenko_electron_2006}%
  \BibitemOpen
  \bibfield  {author} {\bibinfo {author} {\bibfnamefont {D.~V.}\ \bibnamefont
  {Khveshchenko}},\ }\href {\doibase 10.1103/PhysRevLett.97.036802} {\bibfield
  {journal} {\bibinfo  {journal} {Physical Review Letters}\ }\textbf {\bibinfo
  {volume} {97}},\ \bibinfo {pages} {036802} (\bibinfo {year}
  {2006})}\BibitemShut {NoStop}%
\bibitem [{\citenamefont {Onoda}\ \emph {et~al.}(2007)\citenamefont {Onoda},
  \citenamefont {Avishai},\ and\ \citenamefont
  {Nagaosa}}]{onoda_localization_2007}%
  \BibitemOpen
  \bibfield  {author} {\bibinfo {author} {\bibfnamefont {M.}~\bibnamefont
  {Onoda}}, \bibinfo {author} {\bibfnamefont {Y.}~\bibnamefont {Avishai}}, \
  and\ \bibinfo {author} {\bibfnamefont {N.}~\bibnamefont {Nagaosa}},\ }\href
  {\doibase 10.1103/PhysRevLett.98.076802} {\bibfield  {journal} {\bibinfo
  {journal} {Physical Review Letters}\ }\textbf {\bibinfo {volume} {98}},\
  \bibinfo {pages} {076802} (\bibinfo {year} {2007})}\BibitemShut {NoStop}%
\bibitem [{\citenamefont {Zhang}\ \emph
  {et~al.}(2022{\natexlab{b}})\citenamefont {Zhang}, \citenamefont {Wang},
  \citenamefont {Zhang}, \citenamefont {Du},\ and\ \citenamefont
  {Pantelides}}]{zhang_thermal_2022}%
  \BibitemOpen
  \bibfield  {author} {\bibinfo {author} {\bibfnamefont {Y.-T.}\ \bibnamefont
  {Zhang}}, \bibinfo {author} {\bibfnamefont {Y.-P.}\ \bibnamefont {Wang}},
  \bibinfo {author} {\bibfnamefont {Y.-Y.}\ \bibnamefont {Zhang}}, \bibinfo
  {author} {\bibfnamefont {S.}~\bibnamefont {Du}}, \ and\ \bibinfo {author}
  {\bibfnamefont {S.~T.}\ \bibnamefont {Pantelides}},\ }\href {\doibase
  10.1063/5.0089967} {\bibfield  {journal} {\bibinfo  {journal} {Applied
  Physics Letters}\ }\textbf {\bibinfo {volume} {120}},\ \bibinfo {pages}
  {222201} (\bibinfo {year} {2022}{\natexlab{b}})}\BibitemShut {NoStop}%
\bibitem [{\citenamefont {Agarwala}\ and\ \citenamefont
  {Shenoy}(2017)}]{agarwala_topological_2017}%
  \BibitemOpen
  \bibfield  {author} {\bibinfo {author} {\bibfnamefont {A.}~\bibnamefont
  {Agarwala}}\ and\ \bibinfo {author} {\bibfnamefont {V.~B.}\ \bibnamefont
  {Shenoy}},\ }\href {\doibase 10.1103/PhysRevLett.118.236402} {\bibfield
  {journal} {\bibinfo  {journal} {Physical Review Letters}\ }\textbf {\bibinfo
  {volume} {118}},\ \bibinfo {pages} {236402} (\bibinfo {year}
  {2017})}\BibitemShut {NoStop}%
\bibitem [{\citenamefont {Wang}\ \emph {et~al.}(2022)\citenamefont {Wang},
  \citenamefont {Cheng}, \citenamefont {Liu}, \citenamefont {Liu},\ and\
  \citenamefont {Huang}}]{wang_structural_2022}%
  \BibitemOpen
  \bibfield  {author} {\bibinfo {author} {\bibfnamefont {C.}~\bibnamefont
  {Wang}}, \bibinfo {author} {\bibfnamefont {T.}~\bibnamefont {Cheng}},
  \bibinfo {author} {\bibfnamefont {Z.}~\bibnamefont {Liu}}, \bibinfo {author}
  {\bibfnamefont {F.}~\bibnamefont {Liu}}, \ and\ \bibinfo {author}
  {\bibfnamefont {H.}~\bibnamefont {Huang}},\ }\href {\doibase
  10.1103/PhysRevLett.128.056401} {\bibfield  {journal} {\bibinfo  {journal}
  {Physical Review Letters}\ }\textbf {\bibinfo {volume} {128}},\ \bibinfo
  {pages} {056401} (\bibinfo {year} {2022})}\BibitemShut {NoStop}%
\bibitem [{\citenamefont {Guo}\ \emph {et~al.}(2010)\citenamefont {Guo},
  \citenamefont {Rosenberg}, \citenamefont {Refael},\ and\ \citenamefont
  {Franz}}]{guo_topological_2010}%
  \BibitemOpen
  \bibfield  {author} {\bibinfo {author} {\bibfnamefont {H.-M.}\ \bibnamefont
  {Guo}}, \bibinfo {author} {\bibfnamefont {G.}~\bibnamefont {Rosenberg}},
  \bibinfo {author} {\bibfnamefont {G.}~\bibnamefont {Refael}}, \ and\ \bibinfo
  {author} {\bibfnamefont {M.}~\bibnamefont {Franz}},\ }\href {\doibase
  10.1103/PhysRevLett.105.216601} {\bibfield  {journal} {\bibinfo  {journal}
  {Phys. Rev. Lett.}\ }\textbf {\bibinfo {volume} {105}},\ \bibinfo {pages}
  {216601} (\bibinfo {year} {2010})}\BibitemShut {NoStop}%
\bibitem [{\citenamefont {Ying}\ \emph {et~al.}(2016)\citenamefont {Ying},
  \citenamefont {Gu}, \citenamefont {Chen}, \citenamefont {Wang}, \citenamefont
  {Jin}, \citenamefont {Zhao}, \citenamefont {Zhang},\ and\ \citenamefont
  {Chen}}]{ying_anderson_2016}%
  \BibitemOpen
  \bibfield  {author} {\bibinfo {author} {\bibfnamefont {T.}~\bibnamefont
  {Ying}}, \bibinfo {author} {\bibfnamefont {Y.}~\bibnamefont {Gu}}, \bibinfo
  {author} {\bibfnamefont {X.}~\bibnamefont {Chen}}, \bibinfo {author}
  {\bibfnamefont {X.}~\bibnamefont {Wang}}, \bibinfo {author} {\bibfnamefont
  {S.}~\bibnamefont {Jin}}, \bibinfo {author} {\bibfnamefont {L.}~\bibnamefont
  {Zhao}}, \bibinfo {author} {\bibfnamefont {W.}~\bibnamefont {Zhang}}, \ and\
  \bibinfo {author} {\bibfnamefont {X.}~\bibnamefont {Chen}},\ }\href {\doibase
  10.1126/sciadv.1501283} {\bibfield  {journal} {\bibinfo  {journal} {Science
  Advances}\ }\textbf {\bibinfo {volume} {2}},\ \bibinfo {pages} {e1501283}
  (\bibinfo {year} {2016})}\BibitemShut {NoStop}%
\end{thebibliography}%

\end{document}